\documentclass{article}

\usepackage[a4paper, total={6in, 10in}]{geometry}
\usepackage{tabularx} 
\usepackage{graphicx}
\usepackage[svgnames]{xcolor}
\usepackage{amsmath}
\usepackage{amssymb}
\usepackage{hyperref}
\usepackage[utf8]{inputenc}
\usepackage{xspace}
\usepackage{bm,array}
\usepackage[vlined,ruled]{algorithm2e}

\hypersetup
{
	pdftitle={Authorship Verification based on Compression-Models},    
	pdfauthor={Halvani et al.},     
	pdfsubject={Authorship Verification},   
	pdfcreator={},   
	pdfproducer={},  
	pdfkeywords={Authorship Verification} {Compression Models} {Intrinsic Verification}, 	
	pdfpagemode=UseNone,
	pdfstartview={XYZ null null null},
	colorlinks = true,
	linkcolor  = red,	
	citecolor  = blue,
	urlcolor  = blue,
	anchorcolor = blue
}

\newcommand*{\GLADMethod}       {\emph{GLAD Method}\xspace}

\newcommand*{\KoppelMethod}     {\emph{Impostors Method}\xspace}
\newcommand*{\StamatatosMethod} {\emph{Profile-Based Method}\xspace}

\newcommand*{\A}      {\mathcal{A}}
\newcommand*{\Aunk}   {\widetilde{\A}}
\newcommand*{\notA}   {\neg\A}
\newcommand*{\D}      {\mathcal{D}}
\newcommand*{\DA}     {{\D_{\A}}}
\newcommand*{\Dknown} {\D_{\A\vphantom{\Aunk}}}
\newcommand*{\Dunk}   {\D_{\Aunk}}

\newcommand*{\Arefset} {\mathbb{D}_{\A}}

\newcommand*{\panThirteen}     {{\mathcal{C}}_{\textrm{PAN13}}}
\newcommand*{\panFourteenEs}   {{\mathcal{C}}_{\textrm{PAN14es}}}
\newcommand*{\panFourteenNo}   {{\mathcal{C}}_{\textrm{PAN14no}}}
\newcommand*{\panFifteen}      {{\mathcal{C}}_{\textrm{PAN15}}}

\newcommand*{\panThirteenTr}     {{\panThirteen}_{\textrm{Tr}}}
\newcommand*{\panFourteenEsTr}   {{\panFourteenEs}_{\textrm{Tr}}}
\newcommand*{\panFourteenNoTr}   {{\panFourteenNo}_{\textrm{Tr}}}
\newcommand*{\panFifteenTr}      {{\panFifteen}_{\textrm{Tr}}}

\newcommand*{\panThirteenEval}   {{\panThirteen}_{\textrm{Eval}}}
\newcommand*{\panFourteenEsEval} {{\panFourteenEs}_{\textrm{Eval}}}
\newcommand*{\panFourteenNoEval} {{\panFourteenNo}_{\textrm{Eval}}}
\newcommand*{\panFifteenEval}    {{\panFifteen}_{\textrm{Eval}}}

\newcommand*{\Amazon}     {{\mathcal{C}}_{\textrm{Amazon}}}
\newcommand*{\Koppel}     {{\mathcal{C}}_{\textrm{Koppel}}}
\newcommand*{\Reddit}     {{\mathcal{C}}_{\textrm{Reddit}}}

\newcommand*{\AmazonEval}  {{\Amazon}_{\textrm{Eval}}}
\newcommand*{\KoppelEval}  {{\Koppel}_{\textrm{Eval}}}
\newcommand*{\RedditEval}  {{\Reddit}_{\textrm{Eval}}}

\newcommand*{\auccat}  {AUC$\cdot$c@1\xspace}

\newcommand*{\av}   {authorship verification\xspace}

\newcommand*{\AV}   {AV\xspace}

\newcommand*{\CMs}  {CMs\xspace}

\newcommand*{\M}    {\mathcal{M}}

\newcommand*{\AVhalvani}    {\textrm{AV}_{\textrm{we}}}
\newcommand*{\AVstamatatos} {\textrm{AV}_{\textrm{stat}}}
\newcommand*{\AVglad}       {\textrm{AV}_{\textrm{glad}}}
\newcommand*{\AVkoppel}     {\textrm{AV}_{\textrm{kop}}}

\newcommand*{\eg}  {e.\,g.,\mbox{}\xspace}
\newcommand*{\ie}  {i.\,e.\mbox{}\xspace}

\begin{document}
\title{Authorship Verification based on Compression-Models}
\author{Oren Halvani, Christian Winter, Lukas Graner.}

\date{\vspace{-5ex}}
\maketitle

\begin{abstract}
Compression models represent an interesting approach for different classification tasks and have been used widely across many research fields. 
We adapt compression models to the field of \av (\AV), a branch of digital text forensics. 
The task in \AV is to verify if a questioned document and a reference document of a known author are written by the same person. 
We propose an intrinsic \AV method, which yields competitive results compared to a number of current state-of-the-art approaches, 
based on support vector machines or neural networks. However, in contrast to these approaches our method does not make use of machine learning algorithms, 
natural language processing techniques, feature engineering, hyperparameter optimization or external documents 
(a common strategy to transform \AV from a one-class to a multi-class classification problem). 
Instead, the only three key components of our method are a compressing algorithm, a dissimilarity measure and a threshold, 
needed to accept or reject the authorship of the questioned document. 
Due to its compactness, our method performs very fast and can be reimplemented with minimal effort. 
In addition, the method can handle complicated \AV cases where both, the questioned and the reference document, are not related to each other in terms of topic or genre. 
We evaluated our approach against publicly available datasets, which were used in three international \AV competitions. 
Furthermore, we constructed our own corpora, where we evaluated our method against state-of-the-art approaches and achieved, in both cases, promising results. 
\end{abstract}

\section{Introduction}
Authorship analysis is relevant in many kinds of disputes, cases of crime, and acts of manipulation:
The originator of hate mail or blackmail messages has to be identified unequivocally for being sentenced.
Claims of responsibility for attacks and the authenticity of these claims are important aspects in solving that crime.
False insurance claims might not be written by the declared claimant but by another person involved in the fraud.
The originality of a testament and hence the allocation of heritage can be contested.
Online reviews, comments, opinion statements, and fake news by sock puppets are used to influence the public appearance, popularity, or success of products, political missions, etc. Hence, authorship analysis is a highly important field of forensics.

According to Koppel et al. \cite{KoppelFundamentalProblem:2012} \av (\AV) is the \emph{fundamental problem} of authorship analysis.
The goal of \AV is to determine if, given writing examples from an author $\A$, each text in fact was written by $\A$ \cite{SteinAuthorshipVerification:2008}. 
If this is the case, we can also reformulate the problem to decide if, given $\Arefset$ (a set of reference documents written by $\A$), 
a questioned document $\Dunk \notin \Arefset$ is also written by $\A$ \cite{PAN13Stamatatos:2013}. 

Compression Models (\CMs) have been used widely across different domains such as computer science, biology 
or (computational) linguistics, for more than two decades. In recent years, it can be observed on several bibliographic databases and indexes such as \textit{DBLP\footnote{\url{http://dblp.uni-trier.de}}}, \textit{Google Scholar\footnote{\url{https://scholar.google.com}}} or \textit{Microsoft Academic\footnote{\url{https://academic.microsoft.com}}}, that \CMs attract more and more researchers, especially in the field of text classification (\eg \cite{YuvalMartonTC:2005, PereiraCompressionTC:2015, SaikrishnaCompressionTC:2016} just to mention a few). We adapt compression models to \AV which, from a technical point of view, is an instance of text classification problems, 
where the subject is to classify the text regarding its writing style, rather than its content. 

We propose a new intrinsic \AV method based on \CMs, which offers many benefits and contributions.
Our method yields recognition results very similar to a number of current state-of-the-art approaches,
but without relying on sophisticated machine learning techniques (\eg support vector machines or neural networks) or 
natural language processing components (\eg part-of-speech taggers or dependency parsers). 
Instead, our method requires only a compressing algorithm\footnote{Since for almost any popular programming language (\eg C/C++/C\#, Java or Python), existing compression libraries are available, reimplementation is unnecessary.}, a simple dissimilarity measure as well as 
a strategy to determine a threshold that serves as the authorship acceptance criterion. 
In contrast to the majority of existing traditional \AV methods, our approach avoids the entire feature engineering process including the definition, 
extraction and selection of features, which further reduces its complexity. 
Therefore, the proposed method can be reimplemented easily by the community or even by those who are not familiar with the field of \AV. 
Besides the threshold determination (which is obligatory for almost any \AV approach), our method does not generate a model and thus, implies lazy learning.
Moreover, the proposed method features a very fast\footnote{This fact is surprisingly contradictory to the claim of Yuval et al., who state that compression based classification methods suffer from slow running time \cite{YuvalMartonTC:2005}.} runtime (for example, an evaluation of a corpus comprising 500 \av cases 
is processed in few seconds on an off-the-shelf laptop (Intel$\textsuperscript{\textregistered}$ Core$\textsuperscript{\texttrademark}$ i5-3210M processor, 16\,GB RAM). 
Hence, our method is applicable in scenarios (or environments) where runtime matters. 

Another advantage of our method is its ability to solve \AV cases, where the questioned and the reference document 
differ in terms of topic or genre (so-called cross-topic/genre problems) and, furthermore, are short regarding their text lengths. 
Both cases are considered challenging in the field of \AV \cite{BrocardoShortAV:2013, PAN15Stamatatos:2015}.  
In addition to the proposed method, we also offer a number of (cross- and mixed-topic) corpora that enable to reproduce 
our results and can also be used to train and evaluate new \AV methods.

The rest of this paper is structured as follows. 
In Section~\ref{RelatedWork} we review related work and existing approaches that, partially, serve as baselines in the later experiments.   
In Section~\ref{ProposedMethod} we introduce our \AV method, where we explain in detail all necessary steps to reproduce our results.
In Section~\ref{Experiments} we present our experiments based on the proposed \AV method and the introduced baselines, 
and discuss our observations. Finally, we draw our conclusions in Section~\ref{Conclusions} and provide several ideas for future work. 
\section{Related work} \label{RelatedWork}
In the following sub-sections we give an overview of noteworthy \AV approaches, which performed quite well in the context of three PAN\footnote{PAN (\url{http://pan.webis.de})  is a series of scientific events and shared tasks on digital text forensics. Note, the shared task ``Author Identification'' deals with \AV.)} competitions. A part of the introduced \AV methods, marked with ($\ast$) in the sub-section titles, serve as baselines in the later experiments.
Note that some of the introduced approaches model \AV as a binary-class-classification problem.
For reasons of readability, we use the following notation: The target class (the known or reference author) is denoted by $\A$,
while the outlier class (all other possible authors than $\A$) is denoted by $\notA$. The set of documents belonging to $\A$ is
denoted by $\Arefset$, whereas the set of documents belonging to $\notA$ is given by $\mathcal{O}$.
\subsection{Methods by Veenman and Li}

One of the few attempts that adapt \CMs to \av was introduced by Veenman and Li \cite{VeenmanPAN13:2013}.
Overall, the authors proposed three \AV methods based on \CMs, where they transformed the \AV task from a one-class to a two-class classification problem.
More concretely, the authors do not focus on the \textit{recognition}, but instead, on the \textit{discrimination} between $\A$ and $\notA$.

All three methods are \textit{extrinsic}, which means that they rely on external documents (examples of other authors) to model $\notA$.
However, since external documents not exist beforehand ($\mathcal{O} = \emptyset$) Veenman and Li collected external data, where they used
the following procedure: Since they knew that the documents within the training corpus were (approx. 1,000 words long)
fragments of engineering text books, the authors searched for similar books on the web (by using substrings) and gathered 66 books authored by 46 different authors.
After preprocessing, they created 2--75 documents per text book with roughly 1,000 words (6,000--8,000 characters),
which were added to $\mathcal{O}$ \cite{VeenmanPAN13:2013}.

Veenman and Li chose the \textit{Prediction by Partial Matching} (PPMd) algorithm to compress the documents within the training corpus and the
\textit{Compression-based Dissimilarity Measure} (CDM, see Section~\ref{ProposedMethod}) to measure the nearness between the questioned document
$\Dunk$ and the documents in $\Arefset$ and $\mathcal{O}$.

In the first method \cite[Sect.~4.1]{VeenmanPAN13:2013} denoted as \textit{Nearest Neighbor with Compression Distances},
the authors compute the dissimilarity scores between the questioned document $\Dunk$ and all documents in $\Arefset$ and $\mathcal{O}$.
If the document with the smallest dissimilarity score $\D$ belongs to $\Arefset$ then $\Dunk$ is believed to be written by
the reference author $\A$. Otherwise, if $\D$ belongs to $\mathcal{O}$ then $\Dunk$ is believed to be written by some other author $\notA$ \cite{VeenmanPAN13:2013}.

For the second method \cite[Sect.~4.2]{VeenmanPAN13:2013}, the authors used the \textit{Lowest Error in a Sparse Subspace} (LESS) classifier developed by Veenman and Tax \cite{VeenmanTax:2005} in order to discriminate between $\A$ and $\notA$. They select a number of ``prototypes'' from $\mathcal{O}$,
and create a real-valued vector for each document by calculating the CDM between the document and all prototypes.
These vectors are then used as input for the LESS classifier, which is quite similar to a linear classifier with a quadratic kernel.

The third \AV method \cite[Sect.~4.3]{VeenmanPAN13:2013} uses the same classification method as the second one, but the number of documents
in $\Arefset$ is increased artificially by resampling (bootstrapping) new documents from the given reference documents.
This reduces the error rate compared to the second approach. However, the first method performed best in their experiments.

In the later experiments we will provide a differentiation between our method and the first approach of Veenman and Li,
which performed amongst the highest-ranking methods in the PAN 2013 \AV competition \cite{PAN13Stamatatos:2013}.

\subsection{Impostors Method (\texorpdfstring{$\ast$}{*})}
One of the most successful \AV approaches is the well-known \KoppelMethod, proposed by Koppel and Winter \cite{KoppelImpostor:2014}.
The method works in two stages. First, documents needs to be gathered according to a data collection procedure (\eg by using a search engine) that serve as impostors.
Second, feature randomizations are used to iteratively measure the cosine similarity between pairs of documents.
If, given this measure, a suspect is picked out from among the impostor set with sufficient salience, then the suspect is considered
as the author $\A$ of the questioned document $\Dunk$ \cite{KoppelImpostor:2014}. Thus, Koppel and Winter transform the AV problem
(in the same way as Veenman and Li \cite{VeenmanPAN13:2013} do) from a one-class to a multi-class classification
problem ($\A$ against $\notA = $ impostors).


The best-performing \AV approaches of the PAN competitions 2013 and 2014 were based on the \KoppelMethod \cite{PAN13Stamatatos:2013, PAN14Stamatatos:2014}.
But even outside of the PAN competitions the \KoppelMethod gained a lot of attraction by researchers in the \AV field, where it has been reimplemented
and extended in various ways. However, despite of the success of the \KoppelMethod, it also suffers from a number of drawbacks.
For example, it seems to be considerably harder to distinguish same-author and different-author pairs if both differ
in terms of genre and topic \cite{KoppelImpostor:2014}. That might be challenging when generating impostors regarding specific genres or topics
which are not available. Furthermore, since the \KoppelMethod is extrinsic (construction of impostors requires external documents),
it must be guaranteed that impostor documents are not written by the authors of the document pair which are subject of verification \cite{KoppelImpostor:2014}.
Another drawback, highlighted by researchers \cite{KhonjiIraqiAV:2014, SebastianRuderCNNbasedAA:2016} who reimplemented the method,
is the computational expense, which makes the \KoppelMethod unsuitable in scenarios where runtime matters (\eg in web applications).

\subsection{Profile-Based Method (\texorpdfstring{$\ast$}{*})}
The \StamatatosMethod is an intrinsic \AV approach, proposed by Stamatatos and Potha \cite{StamatatosPotha:2014}.
The authors reshaped the original method \cite{KeseljAAProfiles:2003}, which initially was used in the context of authorship attribution, a related field of \AV.
The procedure of the method is quite simple: First, all documents of $\A$ are concatenated into one long document $\DA$, which represents the profile of $\A$.
Next, the $L_k$ most frequent character $n$-grams are extracted from $\DA$. Simultaneously, the $L_u$ most frequent character $n$-grams are extracted from the questioned document $\Dunk$. The generated feature vectors are then compared to each other, given an appropriate dissimilarity function.
If the resulting dissimilarity score exceeds a threshold, learned during training, then $\Dunk$ is believed to be written by $\A$.

The \StamatatosMethod offers a number of advantages that led to the decision why we chose it as a baseline in our experiments.
Regarding the PAN 2013 \AV competition the \StamatatosMethod was able to outperform every single participant in terms of the $F_{1}$
measure\footnote{This measure is explained formally in Section~\ref{PerformanceMeasure}.} \cite{StamatatosPotha:2014}.
Furthermore, it benefits from its compactness, not only from the algorithmic point of view, but also in terms of the involved parameters.
Besides these, the method is quite fast, which makes it applicable in real-time \AV scenarios.


\subsection{GLAD (\texorpdfstring{$\ast$}{*})}
The intrinsic \av system GLAD (Groningen Lightweight Authorship Detection) proposed by Hürlimann et al. \cite{GLAD:2015} uses a compression feature among a diverse set of features. In total, they use 29 features representing seven different kinds of information extracted from the documents. Their compression feature is a modified version of the CDM (see Section~\ref{ProposedMethod}). Compared to the CDM, they use a reciprocal formula and insert the compression ratio instead of the length of compressed documents into that formula. The GLAD system classifies \AV problems with a binary SVM, trained on a set of positive and negative example problems. The classification of a problem itself with the trained SVM is intrinsic, \ie it does not use any additional documents outside the problem. 

The authors evaluated their system regarding the strength of the different feature groups, and their measurements show that their compression feature has only moderate performance \cite[Fig.\,3]{GLAD:2015} and has only a small contribution to the performance of the full feature set \cite[Fig.\,2]{GLAD:2015}. The GLAD system was among the best systems in the PAN 2015 competition except for the English language, where it had moderate performance \cite{PAN15Stamatatos:2015}. However, our experiments reveal that the opposite is the case. Note that Hürlimann et al. made their \AV method publicly available\footnote{\url{https://github.com/pan-webis-de/glad}} and, thus, a reimplementation was not needed beforehand.

\section{Proposed \av method} \label{ProposedMethod}
This section provides a detailed outline of our \AV approach. As a first step we formalize the \AV problem, which is the key subject of our method.
In the next step, we describe how the data is represented. Next, we elaborate how the similarities regarding the documents are computed. Afterwards, we explain our strategy to determine the threshold and, based on it, how the verification (acceptance/rejection of the authorship) is performed.

\paragraph{Formalization of the \AV Problem:} We define a given \AV problem $\rho$ formally as $\rho = (\Dunk, \Arefset)$. Here, $\Dunk$ denotes the questioned document of an author $\Aunk$, who claims (or is suspected) to be $\A$, while $\Arefset = \{\D_{1, \A}, \D_{2, \A},\ldots \}$ denotes the set of sample documents that in fact
are written by the known/reference author $\A$.

\paragraph{Data representation:} We concatenate all sample documents in $\Arefset$ into a single document $\Dknown$, such that the problem
is simplified from $\rho = (\Dunk, \Arefset)$ to $\rho = (\Dknown, \Dunk)$. Next, we make use of a predefined compression algorithm in order to compress
$\Dknown$ to $C(\Dknown)$, $\Dunk$ to $C(\Dunk)$, and the concatenation $\Dknown\Dunk$ to $C(\Dknown\Dunk)$.
In total, we used for our experiments five (already implemented\footnote{For PPMd we used the \textit{SharpCompress} library by Adam Hathcock: \url{https://github.com/adamhathcock/sharpcompress} and for the remaining compressors \textit{SharpZipLib}, developed and maintained by Mike Krüger, John Reilly, David Pierson and Neil McNeight: \url{https://github.com/icsharpcode/SharpZipLib}.}) compressors: PPMd, Gzip, BZip2, Zip and LZW.

\paragraph{Feature engineering:} Since our approach bypass the definition, extraction and selection of features, this step is skipped.

\paragraph{Computing similarities:} Given the compressed files, we need an appropriate measure to determine the nearness between $\Dknown$ and $\Dunk$.
In total, we implemented three dissimilarity measures designed for \CMs, which measure dissimilarity in terms of the length of compressed documents. The length of a compressed document $C(x)$ is -- for convenience and in accordance to the cited publications -- also denoted as $C(x)$. The dissimilarity measures were also used in plenty of previous research works.

The first measure, the \textit{Normalized Compression Distance} (NCD) developed by Cilibrasi, Li et al. \cite{SimNCDCilibrasi:2005, SimNCDLi:2004}, is defined as:
\begin{equation}
\mathrm{NCD}(x,y) = \frac{C(xy) - \min\{C(x),C(y)\}}{\max\{C(x),C(y)\}}.
\end{equation}
This dissimilarity measure is derived from the universal Normalized Information Distance (NID) \cite{SimNCDLi:2004}, which is defined in terms of the Kolmogorov complexity, and hence the NCD has the most elaborated theoretical foundation among the considered dissimilarity measures. If a compressor is a \textit{normal}, as defined by Cilibrasi and Vitányi \cite[Definition~3.1]{SimNCDCilibrasi:2005}, the NCD maps into the interval $[0;1]$ and satisfies the axioms of a metric up to small deviations caused by the imperfections (allowed errors) in the normality of the compressor \cite[Theorem~6.2]{SimNCDCilibrasi:2005}. Cilibrasi and Vitányi state that the NCD maps ``in practice'' into the interval $[0;1+\varepsilon]$ with $\varepsilon \approx 0.1$ or less \cite[Remark~6.2]{SimNCDCilibrasi:2005}.

The second measure, the \textit{Compression-based Cosine} (CBC) proposed by Sculley and Brodley \cite{SimCosSSculley:2006}, is defined as:
\begin{equation}
\mathrm{CBC}(x,y) = 1 - \frac{C(x) + C(y) - C(xy)}{\sqrt{C(x)C(y)}}.
\end{equation}
This dissimilarity function maps into the interval $[0;1]$, but it is not a metric because it violates the triangle inequality.

The third measure, called \textit{Chen-Li metric} (CLM) by Sculley and Brodley \cite{SimCosSSculley:2006} crediting it to Chen, Li et al. \cite{SimCLMChen:1999, SimCLMLi:2001}, is defined as:
\begin{equation}
\mathrm{CLM}(x,y) = 1 - \frac{C(x) - C(x|y)}{C(xy)}.
\end{equation}
The term $C(x|y)$ is the (length of the) conditional compression of $x$ given $y$ as auxiliary input for compression and decompression. Standard implementations of compression algorithms must be enhanced for using $y$ as auxiliary input, but according to Sculley and Brodley \cite{SimCosSSculley:2006}, the approximation $C(x|y) \approx C(xy) - C(y)$ can be used in order to calculate the CLM with ``off-the-shelf compressors''. Cilibrasi and Vitányi even define $C(x|y) = C(xy) - C(y)$ since they do not consider compressors with the capability of conditional compression \cite[Definition~3.2]{SimNCDCilibrasi:2005}. The CLM maps into the interval $[0;1]$. Furthermore, it is a metric except for minor errors. This is proven by Li et al. for compression in terms of the Kolmogorov complexity \cite[Theorem~1]{SimCLMLi:2001}, but their proof can be extended to compression with normal compressors as defined by Cilibrasi and Vitányi \cite[Definition~3.1]{SimNCDCilibrasi:2005}.

Many authors also use the \emph{Compression-based Dissimilarity Measure} (CDM) proposed by Keogh et al. \cite{Keogh:2004} due to the simplicity of the formula:
\begin{equation}
\mathrm{CDM}(x,y) = \frac{C(xy)}{C(x) + C(y)}.
\end{equation}
This dissimilarity score lies in the interval $[0.5;1]$, and it is related to the (approximated) CLM by the order-preserving mapping
\begin{equation}
\textrm{CLM} = 2 - \frac{1}{\textrm{CDM}}. \label{eq:CDM-CLM-equiv}
\end{equation}
Thus, both measures are equivalent for our purpose. We decided to work with the CLM because it ranges from 0 to 1.

All four dissimilarity measures are closely related, as shown by Sculley and Brodley \cite{SimCosSSculley:2006}: Each measure can be written in the form
\begin{equation}
S(x,y) = 1 - \frac{C(x) + C(y) - C(xy)}{N(x,y)},
\end{equation}
where $N(x,y)$ is a normalization term \cite[Table~1 in Sect.~3.2]{SimCosSSculley:2006}. Based on this, any of the dissimilarity scores will exceed 1 if and only if $C(x) + C(y) - C(xy) < 0$. This can happen due to imperfections in the normality of the compressor. The normalization terms of the dissimilarity measures obey the following relations:
\begin{equation}
\begin{alignedat}{2}
&& N_\mathrm{CBC}(x,y) &= \sqrt{C(x)C(y)} \\
& \leq {} & N_\mathrm{NCD}(x,y) &= \max\{C(x),C(y)\} \\
& \leq {} & N_\mathrm{CLM}(x,y) &= C(xy) \\
& \leq {} & N_\mathrm{CDM}(x,y) &= C(x) + C(y).
\end{alignedat}
\end{equation}
Note that $C(xy)$ can violate the inequalities due to imperfections of the compressor. Any dissimilarity score can also fall below the respective lower bound of its nominal range, which is 0.5 for the CDM and 0 for the other measures, due to imperfections of the compressor. This happens most likely for the CBC, as it has the smallest normalization factor. The CDM falls below 0.5 if and only if the CLM falls below 0 due to the equivalence observed above \eqref{eq:CDM-CLM-equiv}. We wish to highlight that in our experiments we did not observe any case, where a score determined with one of the used measures was outside the interval $[0;1]$.

\paragraph{Threshold determination:} Once we computed the dissimilarity $\M(\Dknown, \Dunk)$ with a fixed measure $\M \in \{ \mathrm{NCD}, \mathrm{CBC}, \mathrm{CLM} \}$,
we need a threshold $\theta$ to judge whether $\Dknown$ and $\Dunk$ were written by the same author $\A$ or not.
Our strategy to define $\theta$ requires a training corpus consisting of $n$ problems equally distributed regarding true (\texttt{Y}) and false (\texttt{N}) authorships.
Given the chosen measure, we compute for each problem $\rho$ in this corpus a dissimilarity score $s_\rho$.
Then, we determine $\theta$ based on the EER (\textit{equal error rate}), \ie we select the threshold where the false acceptance rate and the false rejection rate are equal. The detailed procedure to compute $\theta$ is listed in Algorithm~\ref{DetermineThresholdEER}.

\begin{algorithm}
	\SetKwInput{KwData}{Input}
	\SetKwInput{KwResult}{Output}
	\SetKw{KwBreak}{break}
	\SetKw{KwContinue}{continue}
	\SetKw{KwThrow}{throw}
	
	\KwData{$Y$ (the dissimilarity scores of the \texttt{Y} problems), $N$  (the dissimilarity scores of the \texttt{N} problems)}
	\KwResult{$\theta$ (the threshold)}
	\BlankLine

	\If{$\mathrm{length}(Y) \neq \mathrm{length}(N)$}
	{
		Exception $\leftarrow$ "Number of Y and N problems mismatch!"\;
		\KwThrow Exception\;
	}
	\BlankLine
	
	Sort $Y$ and $N$ in ascending order\;
	$\ell \leftarrow \mathrm{length}(Y)$\;
	$i \leftarrow 0$\;
	$j \leftarrow \ell - 1$\;
	\BlankLine

	\For{$k \leftarrow 0$ \KwTo $\ell - 1$}
	{
		\If{$Y_{i} < N_{j}$}
		{
			$i \leftarrow i + 1$\;
			$j \leftarrow j - 1$\;
			\KwContinue\;
		}
		\If{$Y_{i} = N_{j}$}
		{
			$\theta \leftarrow Y_{i}$\;
			\KwBreak\;
		}
		\If{$Y_{i} > N_{j}$}
		{
			\uIf{$i = 0$}
			{
				$\theta \leftarrow \frac{1}{2} (Y_{i} + N_{j})$\;
			}
			\Else
			{
				$\theta \leftarrow \frac{1}{2} \big( \min(Y_{i}, N_{j + 1}) + \max(Y_{i - 1}, N_{j}) \big)$\;
			}
			\KwBreak\;
		}
	}
	\If{$i = \ell$}
	{
		$\theta \leftarrow \frac{1}{2} (Y_{i - 1} + N_{j + 1})$\;
	}
	\BlankLine
	
	\KwRet{$\theta$}

	\caption{Determine threshold via EER.}
	\label{DetermineThresholdEER}
\end{algorithm}
The training procedure is done once beforehand such that $\theta$ is readily available when deciding about the authorship in $\rho$. Note that our approach only requires a training corpus to determine $\theta$. However, the \AV method itself is intrinsic, which means that it is based exclusively on all documents within $\rho$ rather than on external available texts.

\paragraph{Classification:} The last step is to verify the authorship for the problem $\rho$ based on the dissimilarity score
$s_\rho = \M(\Dknown, \Dunk)$ and the decision threshold $\theta$:
\begin{equation}
\textrm{decision}(\rho) = \begin{cases}
\texttt{Y}\ (\textrm{Yes}) & \textrm{if}\ s_\rho < \theta, \\
\texttt{N}\ (\textrm{No})  & \textrm{if}\ s_\rho \geq \theta, \\
\end{cases}
\end{equation}

Another possibility would be to define a specific range $R = [\theta - \varepsilon, \theta + \varepsilon]$,
such that if $s_\rho \in R$ holds, we output an \texttt{unanswered} decision (rather than \texttt{Y}/\texttt{N}), due to uncertainty.
However, this would require additional training in order to learn a stable $R$ across several corpora. Therefore, we leave this option open for future work.
\section{Experiments} \label{Experiments}
In the following sub-sections we describe a number of experiments. First, we explain which corpora we used, how they were generated and which characteristics they cover. Next, we explain how each corpus was preprocessed especially in terms of deduplication and noise removal. Finally, we present our experiments, where the first one focuses on how to find the optimal compression algorithm as well as the most suitable dissimilarity measure. In the second experiment, we build on the findings of the first experiment and measure how well our \AV method performs on the PAN corpora collection. Here, we compare our results with the official results of all participants from the three PAN competitions 2013--2015, given in \cite{PAN13Stamatatos:2013, PAN14Stamatatos:2014, PAN15Stamatatos:2015}. In the last experiment, we measure our \AV method regarding three additional corpora, where we compare our method against the re-implemented baselines, mentioned in Section~\ref{RelatedWork}). 

\subsection{Corpora} \label{Corpora}
Corpora are the key ingredient when it comes to train or measure the quality of \AV methods. In the context of our experiments we made use of seven corpora (splitted into training and evaluation subcorpora). In the following sub-sections we give a brief introduction regarding these corpora, covering their origin, characterization as well as and a compact statistic. All corpora are listed in Table~\ref{tab:AllCorpora}, where the second and third columns denote the number of problems within the training and evaluation subcorpora. Note that regarding our self-compiled corpora\footnote{Available under \url{http://tiny.cc/WSDF2017}} (the last three rows in Table~\ref{tab:AllCorpora}) we splitted the initial corpus into training (20\%) and evaluation (80\%) subcorpora. Since the PAN corpora (the first four rows in Table~\ref{tab:AllCorpora}) were already separated into training and evaluation subcorpora, additional splitting was not needed. Note that all given statistics (number of problems, averaged document lengths, etc.) in the following sub-sections refer to the evaluation corpora.
\begin{table}
	\centering
	\resizebox{\columnwidth}{!}{
		\begin{tabular}{|l|r|r|r|r|r|} 
			\hline
		    \textbf{Corpus} $\bm{\mathcal{C}}$  & $\bm{|\mathcal{C}_{\textrm{Train}}|}$ & $\bm{|\mathcal{C}_{\textrm{Eval}}|}$ & \textbf{Genre} & \textbf{Topic} & \textbf{Source} \\ \hline
			$\panThirteen$   & 10  & 30    & textbooks          & same topic   & \cite{PAN13Stamatatos:2013} \\ \hline
			$\panFourteenEs$ & 200 & 200   & essays             & cross-topic  & \cite{PAN14Stamatatos:2014} \\ \hline
			$\panFourteenNo$ & 100 & 200   & novels             & cross-topic  & \cite{PAN14Stamatatos:2014} \\ \hline
			$\panFifteen$    & 100 & 500   & dialog lines       & cross-topic  & \cite{PAN15Stamatatos:2015} \\ \hline			
			$\Reddit$        & 200 & 800   & social news        & cross-topic  & self-compiled \\ \hline			
			$\Koppel$        & 400 & 1,600 & blog posts         & mixed topics & modified, \cite{KoppelSchlerArgamonPennebaker:2006} \\ \hline
			$\Amazon$        & 646 & 2,582 & product reviews    & cross-topic  & modified, \cite{AmazonReviewCorpus:2015} \\ \hline
		\end{tabular}
	}
	\caption{Statistics and references regarding all corpora.} \label{tab:AllCorpora}
\end{table}

\subsubsection{PAN Corpora Collection} 
The PAN corpora\footnote{Available under \url{http://pan.webis.de}} are (to our best knowledge) the most prominent corpora in the field of \AV. For our purpose, we used all English corpora from 2013--2015 \cite{PAN13Stamatatos:2013, PAN14Stamatatos:2014, PAN15Stamatatos:2015}, which we denote as $\panThirteen, \panFourteenEs, \panFourteenNo$ and $\panFifteen$. 

The $\panThirteen$ corpus is predominantly compiled from textbooks from computer science and related disciplines. 157 documents, of which 97 are unique, were collected from 16 different authors. The corpus compromises 30 problems (in average five known documents per problem), one of which is unknown, documents sized at $\approx$ 7 KByte \cite{PAN13Stamatatos:2013}. While some of these documents are very narrow regarding their genre, others are more divergent in their content. This aims to prevent authorship detection by a simple content detection and, therefore, makes the verification task more challenging. 

The $\panFourteenEs$ corpus was derived from the existing \textit{The Uppsala Student English (USE) corpus}. Each document represents an essay of $\approx$ 5 KByte and falls in one of the following types of writing style: personal, formal, academic. The problems then were constructed so that all documents in one problem were of the same style and their authors were similar in age. In total, 200 problems (each with approximately three known documents) were constructed from 718 documents (625 unique) of 435 authors \cite{PAN14Stamatatos:2014}.

The $\panFourteenNo$ specifies on both a narrow writing style and content. The corpus is focusing on a very small subgenre of speculative and horror fiction, namely ``Cthulhu Mythos''. The special genre, which is based on writings of the American H.P. Lovecraft, is characteristic of florid prose and a vocabulary very unusual for English literature.  43 unique documents were collected from different online sources, including Project Gutenberg and FanFiction. Each problem consist solely of one known and one unknown document. In total, the corpus comprises 200 problems and covers document sizes from 9 to 68 KByte \cite{PAN14Stamatatos:2014}.

The $\panFifteen$ corpus includes dialog lines from plays, excluding named entities like speaker names or lists of characters. 
According to \cite{PAN15Stamatatos:2015}, the corpus is considered to be challenging for at least three reasons. 
First, for each problem in the corpus only one known document per problem is provided. Second, the documents are quite short (on average 536 words per document). 
Third, the known and unknown document differ in terms of topic. In total, the $\panFifteen$ corpus comprises 500 problems covering 1,000 documents, where 83 are unique with a size of $\approx$ 2 KByte. 

\subsubsection{Koppel Blogs AV Corpus}
The \emph{Koppel Blogs AV-Corpus} is a derivate of the existing \emph{The Blog Authorship Corpus}, released by Schler et al. \cite{KoppelSchlerArgamonPennebaker:2006}. In its original form, the corpus comprises postings of 19,320 bloggers, gathered from \emph{blogger.com} in August 2004 and incorporates a total of 681,288 posts (in total over 140 million words). Since the corpus was designed for author profiling (a related discipline to \AV) there was a need to apply a transformation regarding the structure, in order to be used in the context of AV. For this, we extracted postings of 2,000 authors (or precisely users, as several users might refer to the same person) and aggregated them in such a way, that for each user there are four documents with an average file size of $\approx$ 4 KByte. 
	
\subsubsection{Amazon Product Data AV Corpus}
The \emph{Amazon Product Data AV Corpus} is also a derivate of an existing corpus, released by McAuley et al. \cite{AmazonReviewCorpus:2015}. The original corpus contains product reviews and metadata from \emph{Amazon}, including 142.8 million reviews, gathered between 1996--2014. More precisely, the corpus includes reviews (ratings, text, helpfulness votes), product metadata (descriptions, category information, price, brand, and image features) as well as links (also viewed/also bought graphs). However, for the purpose of \av we only used of the reviewer IDs and their associated reviews. In total, the \emph{Amazon Product Data AV Corpus} comprises 3,228 problems (21,534 documents), where for each problem there are $\approx$ 6 known documents ($\approx$ 4 KByte file size  per document). In order to enable a cross-topic property, we constructed the corpus in such a way that all documents within a problem differ from each other in terms of topics. Here, each topic refers to a specific \textit{Amazon} product category. In total, 17 product categories span the entire corpus such as automotive, beauty or musical instruments. 
	
\subsubsection{Reddit Cross-Topic AV Corpus}
In contrast to the former two this corpus was entirely compiled by us and represents the only corpus that was not derived from an existing corpus. 
It consists of comments written between 2010--2016 from 1,000 Reddit users and thus, comprises 1,000 problems. 
Each problem includes one unknown and four known documents with $\approx$ 7 KByte per document, 
where each document represents an aggregation of reviews coined from the same so-called \emph{subreddit}. 
However, all documents within a problem are disjunct to each other in terms of \emph{subreddits}. 
This was done in the same way to the \emph{Amazon Product Data AV Corpus} in order to enable a cross-topic property for the corpus. 
All subreddits cover exactly 1,388 different topics such as books, news, gaming, music, movies, etc.

\subsection{Preprocessing}
During the construction of our three additional corpora, a number of preprocessing steps were necessary, before they could be used in the later experiments. First, we performed deduplication for each corpus in terms of exact and near-duplicates. To identify near-duplicates, we applied on each pair of texts, or more precisely on their tokenized representation $X$ and $Y$, the overlap coefficient similarity measure:
\begin{equation}
\textrm{overlap}(X, Y) = \frac{|X \cap Y|}{\min(|X|, |Y|)}
\end{equation}
Two texts were considered as near-duplicates if their resulting overlap coefficient exceeded 0.25. We choose this value since our empirical observations have shown that two documents with an overlap coefficient below 0.25 are partly similar in terms of function words, but not in terms of content words. After the deduplication process, we removed noise from the remaining documents in the following manner: First, all available texts were converted to \mbox{UTF-8} format. Next, we removed URLs, HTML tags, encoded HTML strings, \mbox{UTF-8} control characters, consecutive symbols and formulas from the texts. As a last step, we replaced newlines, tabs and consecutive whitespace characters by a single space, such that each document resulted in a single-lined string.

\subsection{Performance Measures} \label{PerformanceMeasure} 
During the past decades, a number of performance measures have become established in the field of \av. Perhaps the simplest one used in \AV (\eg \cite{KoppelOneClassAV:2004,BrennanGreenstadt:2009}) is the \textit{Accuracy} measure, defined as:
\[
   \textrm{Accuracy} = \frac{\textrm{Number of correct answers}}{\textrm{Total number of problems}}
\]

Another (more sophisticated) performance measure is the ROC-AUC\footnote{For reasons of readability, ROC-AUC is often shortened to AUC in the literature.} (\textit{\textbf{A}rea \textbf{U}nder the ROC \textbf{C}urve}), which was used, for instance, by Stamatatos and Potha in their work \cite{StamatatosPotha:2014}. One significant characteristic of AUC is that it is not bounded to a fixed threshold, which is the case of the accuracy measure. This allows a more reliable comparison among different \AV methods. 

Another measure, typically used in information retrieval for binary classification tasks, is $\textrm{F}_1$, defined as follows:
\[ \textrm{F}_{1} = \frac{2 \cdot \textrm{Recall} \cdot \textrm{Precision}} {\textrm{Recall} + \textrm{Precision}} \]

In the PAN-2013 competition \cite{PAN13Stamatatos:2013} the $\textrm{F}_1$ measure was used to evaluate the \av methods of all participants, but non-standard definitions for recall and precision were used:
\[ \textrm{Recall} = \frac{\textrm{Number of correct answers}} {\textrm{Total number of problems}} \]

\[ \textrm{Precision} = \frac{\textrm{Number of correct answers}} {\textrm{Total number of answers}} \]

In recent years, a new performance measure is becoming more and more popular in the AV community, the so-called \auccat that was first introduced in \cite{PAN14Stamatatos:2014}. It represents the product of the already mentioned AUC and the c@1, which is defined as:
\[
   \textrm{c@1} = \frac{1}{n} \biggl( n_c + \left(\frac{n_u \cdot n_c}{n}\right) \biggr)
\]

Here, $n$, $n_c$ and $n_u$ denote the number of problems, the number of correct answers and the number of unanswered problems, respectively. Note that c@1 equals the accuracy measure for the case that an \AV method classifies all problems either as \texttt{Y} or \texttt{N}. If, on the other hand, all problems are left unanswered, then c@1 will be zero. However, c@1 rewards such \AV methods that maintain the same number of correct answers and decrease the number of wrong answers by leaving some problems unanswered \cite{PAN14Stamatatos:2014}. At this point we would like to underline that our \AV method does not benefit from c@1, since (at least in its current form) it does not leave any problem unanswered. 

In order to provide the reader a better comparison regarding our approach, we will use all mentioned measures from above in our experiments. 
In order to avoid redundancy, we omit the accuracy measure as it is identical to c@1, at least regarding our method.

\subsection{Experiment 1: Determining the optimal compressor/dissimilarity measure}
The goal of the first experiment is to find the optimal combination of a compression algorithm $c(\cdot) \in \{ \mathrm{PPMd}, \mathrm{GZip}, \mathrm{BZip2}, \mathrm{Zip}, \mathrm{LZW} \}$ and a dissimilarity measure $\M \in \{ \mathrm{NCD}, \mathrm{CBC}, \mathrm{CLM} \}$ that lead to the highest performance, in terms of the AUC. 
For this, we applied our method on the four PAN training corpora, which results in $60$ different runs. The results regarding these runs are given in Table~\ref{tab:TrainingCorpusMetricCompressor}, where it can be seen in the last row (showing the average AUC among all corpora) that PPMd was able to outperform all other involved compressors, independently of the underlying dissimilarity measure. Moreover, it can be seen in the same row that across all compressors the highest AUC is constantly achieved by the CBC measure. Therefore, we used $c(\cdot)$ = PPMd and $\M$ = CBC for the further experiments. One interesting observation (which, due to a lack of space, did not fit in Table~\ref{tab:TrainingCorpusMetricCompressor}) is the fact, that Zip is by far the fastest among the five compressors. More precisely, Zip is at least twice as fast as the best performing compressor PPMd. 

\begin{table*}
	\small
	\centering
	\resizebox{\textwidth}{!}{
	\begin{tabular}{|l||r|r|r||r|r|r||r|r|r||r|r|r||r|r|r|}
		\hline
		& \multicolumn{3}{c||}{\textbf{PPMd}} & \multicolumn{3}{c||}{\textbf{Gzip}}  & \multicolumn{3}{c||}{\textbf{BZip2}}  & \multicolumn{3}{c||}{\textbf{Zip}}  &\multicolumn{3}{c|}{\textbf{LZW}}  \\ \hline
		Corpus              & NCD & CBC & CLM & NCD & CBC & CLM  & NCD & CBC & CLM & NCD & CBC & CLM & NCD & CBC & CLM \\ \hline
		$\panThirteenTr$   & 0.800 & 1.000 & 1.000 & 0.800 & 1.000 & 0.880 & 0.800 & 1.000 & 0.920 & 0.720 & 0.920 & 0.800  & 0.680 & 0.880 & 0.720 \\
		$\panFourteenEsTr$ & 0.541 & 0.545 & 0.545 & 0.566 & 0.587 & 0.576 & 0.541 & 0.527 & 0.540 & 0.549 & 0.547 & 0.552 	& 0.552 & 0.517 & 0.541 \\
		$\panFourteenNoTr$ & 0.596 & 0.705 & 0.634 & 0.584 & 0.652 & 0.624 & 0.581 & 0.656 & 0.611 & 0.534 & 0.549 & 0.557 	& 0.518 & 0.592 & 0.573 \\
		$\panFifteenTr$    & 0.666 & 0.702 & 0.700 & 0.632 & 0.664 & 0.658 & 0.613 & 0.666 & 0.662 & 0.589 & 0.612 & 0.608 & 0.564 & 0.620 & 0.615 \\ \hline
		Average            & 0.651 & \textbf{0.738} & 0.720 & 0.645 & \textbf{0.726} & 0.685 & 0.634 & \textbf{0.712} & 0.683 & 0.598 & \textbf{0.657} & 0.629 & 0.578 & \textbf{0.652} & 0.612 \\ \hline
	\end{tabular}
}
\caption{Results showing (in terms of AUC) how well the compressing algorithms performed regarding the dissimilarity measures on the PAN training corpora.}
	\label{tab:TrainingCorpusMetricCompressor}
\end{table*}

\subsection{Experiment 2: Evaluation (PAN competition)} 
The findings in Experiment~1 have shown that the PPMd compressor, together with the CBC metric, outperformed all other (compressor/metric)-combinations. Given these both, we applied our \AV method on each of the four PAN training corpora in order to determine a threshold, based on Algorithm~\ref{DetermineThresholdEER}. Each threshold was then used to test our method on the corresponding four PAN evaluation corpora. In the following, we discuss the results given in Tables~\ref{tab:PAN2015Eval}--\ref{tab:PAN2013EvalAUC}.

\paragraph{Results for $\panFifteenEval$:} As can be seen from Table~\ref{tab:PAN2015Eval}, our method outperforms (with the exception of the winning approach of Bagnall) all approaches of the participating teams , the three baselines \textit{PAN13-BASELINE}, \textit{PAN14-BASELINE-1} and \textit{PAN14-BASELINE-2}, as well as the meta-model \textit{PAN15-ENSEMBLE}, which fuses the output of all submitted \AV methods. Note that Bagnall's sophisticated approach \cite{BagnallRNN:2015}, which is based on multi-headed recurrent neural networks, is included within the meta-model. It is noteworthy to underline the fact, that our method yields results very similar to Bagnall's method but, however, in much less time (7 seconds instead of more than 21 hours). To learn more about the performance of our method, we inspected the verification results on the problem level. Here, we observed that our method successfully solved many \AV problems with small document sizes (for example 1.56 KByte) regarding $\Dknown$ and $\Dunk$. In addition to this, we want to stress that the documents within each problem differ in terms of topic, such that the verification task is even more challenging, as it already is. 

\paragraph{Results for $\panFourteenEsEval$:} The results regarding $\panFourteenEsEval$ are given in Table~\ref{tab:PAN2014esEval}, where it, unfortunately, can be seen that our method achieves only moderate results (one rank above the baseline). In order to understand what causes the low recognition performance, we applied a deeper investigation on the $\panFourteenEsEval$ corpus. First, we compared the document lengths regarding the known and unknown document $\Dknown$ and $\Dunk$. For the four shortest texts, we noted document sizes between 2--3 KByte which, as shown in the results for $\panFifteenEval$, seems not to be the primary challenge for the low performance. Therefore, we inspected the common vocabulary between $\Dknown$ and $\Dunk$ for each problem. But it turned out that, on average, all $(\Dknown, \Dunk)$ pairs share a sufficient amount of features (\eg function words or character bigrams) that allow the problems to be successfully solved. Therefore, we leave this challenge for future work.  

\paragraph{Results for $\panFourteenNoEval$:} In contrast to $\panFourteenEsEval$, our method seems to perform much better on the $\panFourteenNoEval$ corpus. From the  results, shown in Table~\ref{tab:PAN2014noEval}, one can see that our approach is ranked among the three best-performing methods. One of these three is the approach of Khonji\&Iraqi \cite{KhonjiIraqiAV:2014}. Their approach, a modified version of the state-of-the-art \KoppelMethod, performs better than our method in terms of the AUC, but at the expense of runtime. 

\paragraph{Results for $\panThirteenEval$:} For the $\panThirteenEval$ evaluation corpus two tables are provided. Table~\ref{tab:PAN2013EvalAccuracy} and Table~\ref{tab:PAN2013EvalAUC} list the evaluation results in terms of $F_{1}$ and AUC, respectively. The rationale behind this is, that in the initial \AV competition PAN 2013, the participants were requested to provide binary scores, in the form of \texttt{Y}/\texttt{N}, which were mandatory to compute the $F_{1}$ and, optional, real numbers in the interval $[0 ;1]$ in order to compute the AUC. However, since not all participants provided real numbers there was a need for two tables. 

As one can see from Table~\ref{tab:PAN2013EvalAccuracy} our approach shares the highest rank together with the methods of Veenman\&Li \cite{VeenmanPAN13:2013} and Seidman \cite{SeidmanPAN13:2013}, in terms of $F_{1}$. When we look on Table~\ref{tab:PAN2013EvalAUC} we can see that Seidman's approach (which, analogous to Khonji\&Iraqi \cite{KhonjiIraqiAV:2014}, is also based on \KoppelMethod) performs somewhat worse than our method, in terms of AUC. 

Another observation, which cannot be seen in both tables is, that the consumed runtime of Seidman's method is very high. Note that we cannot tell the runtime for the $\panThirteenEval$ corpus, as the organizers only provided the overall runtime across three corpora \cite[Table~1]{PAN13Stamatatos:2013}, which in case of Seidman's method is $\approx$ 18 hours. If we naively divide the runtime by 3 (due to three corpora), Seidman's method will still need six hours for the $\panThirteenEval$  corpus, whereas our method can process the same corpus in less than a second. 

Unfortunately, Veenman and Li did not provide AUC scores for their submitted \AV method, such that a direct comparison to our method is not possible. Therefore, we decided at least to provide a differentiation (Table~\ref{tab:VeenmanVsHalvani}) regarding the characteristics between our and their method. 
\begin{table}
	\centering
	\resizebox{\columnwidth}{!}{
		\begin{tabular}{|l|l|l|l|l|l|} 
			\hline
			             & \textbf{\AV type} & \textbf{Threshold}     & \textbf{Dis. Measure}  & \textbf{Compressor} & \textbf{Output} \\ \hline
			Our approach & intrinsic             & Yes, via EER  & CBC            & PPMd       & real score + \texttt{Y}/\texttt{N}\\ \hline
			Veenman\&Li  & extrinsic             & Not provided  & CDM            & PPMd       & only \texttt{Y}/\texttt{N} \\ \hline		
		\end{tabular}
	}
	\caption{Our approach vs. Veenman and Li's method \cite{VeenmanPAN13:2013}.} \label{tab:VeenmanVsHalvani}
\end{table}

\begin{table}
	\centering
	\scalebox{0.84}{
		\begin{tabular}{|l|r|r|r|r|r|}
			\hline
			\bf Team & \bf FS & \bf AUC & \bf c@1  & \bf UP  & \bf Runtime \\ \hline
			Bagnall & 0.614 & 0.811  & 0.757 &  3  & 21:44:03  \\\hline
			\textbf{Our approach} & \textbf{0.605} & \textbf{0.802} & \textbf{0.754} & \textbf{0} & \textbf{00:00:07} \\\hline
			Castro-Castro et al. & 0.520 & 0.750  & 0.694  & 0 &  02:07:20 \\
			Gutierrez et al.  & 0.513  & 0.739  & 0.694  & 39  & 00:37:06 \\
			Kocher and Savoy  & 0.508 &  0.738 &  0.689  & 94 &  00:00:24 \\
			{\color{gray} PAN15-ENSEMBLE}  & {\color{gray} 0.468}  & {\color{gray} 0.786} &  {\color{gray} 0.596} &  {\color{gray} 0} & {\color{gray} -}  \\
			Halvani  & 0.458  & 0.762  & 0.601  & 25 &  00:00:21 \\
			Moreau et al.  & 0.453 &  0.709 &  0.638  & 0 &  24:39:22 \\
			Pacheco et al.  & 0.438  & 0.763  & 0.574  & 2  & 00:15:01 \\
			H{\"u}rlimann et al.  & 0.412 &  0.648  & 0.636 &  5 &  00:01:46 \\
			{\color{gray} PAN14-BASELINE-2} & {\color{gray} 0.409} & {\color{gray}  0.639} & {\color{gray}  0.640}  &{\color{gray} 0} & {\color{gray} 00:26:19} \\
			{\color{gray} PAN13-BASELINE}  & {\color{gray} 0.404} & {\color{gray}  0.654} & {\color{gray}  0.618} & {\color{gray}  0} & {\color{gray}  00:02:44} \\
			Posadas-Dur$\acute{\mathrm{a}}$n et al.  & 0.400  & 0.680  & 0.588 &  0 &  01:41:50 \\
			Maitra et al.  & 0.347  & 0.602 &  0.577 &  10  & 15:19:13 \\
			Bartoli et al.  & 0.323 & 0.578 & 0.559  & 3 &  00:20:33 \\
			G$\acute{\mathrm{o}}$mez-Adorno et al.  & 0.281  & 0.530  & 0.530  & 0  & 07:36:58 \\
			Sol$\acute{\mathrm{o}}$rzano et al.  & 0.259  & 0.517  & 0.500  & 0  & 00:29:48 \\
			Nikolov et al.  &  0.258  & 0.493  & 0.524  & 16  & 00:01:36 \\
			Pimas et al.   & 0.257  & 0.507  & 0.506 &  0  & 00:07:22 \\
			{\color{gray} PAN14-BASELINE-1}  & {\color{gray} 0.249} & {\color{gray}  0.537}  & {\color{gray} 0.464}  & {\color{gray} 159}  &{\color{gray}  00:01:11} \\
			Mechti et al.  & 0.247 &  0.489 &  0.506 &  0 &  00:04:59 \\
			Sari and Stevenson  & 0.201  & 0.401  & 0.500  & 0 &  00:05:47 \\
			Vartapetiance and G. & 0.000  & 0.500  & 0.000  & 500 & - \\ \hline
	\end{tabular}}
	\caption{PAN 2015 results on the English evaluation corpus. The final score FS denotes \auccat and UP the number of unanswered problems. Table adapted from \cite{PAN15Stamatatos:2015}.}
	\label{tab:PAN2015Eval}
\end{table}

\begin{table}
	\centering
	\scalebox{0.84}{
		\begin{tabular}{|l|r|r|r|r|r|}
			\hline
			\bf Team & \bf FS & \bf AUC & \bf c@1  & \bf UP  & \bf Runtime \\ \hline
			{\color{gray} META-CLASSIFIER}  & {\color{gray} 0.531}  & {\color{gray} 0.781} & {\color{gray} 0.680} & {\color{gray} 0} & {\color{gray} ---} \\
			Frery et al. & 0.513 & 0.723 & 0.710 & 15 & 00:00:54 \\
			Satyam et al. & 0.459 & 0.699 & 0.657 & 2 & 00:16:23 \\
			Moreau et al. & 0.372 & 0.620 & 0.600 & 0 & 00:28:15 \\
			Layton & 0.363 & 0.595 & 0.610 & 0 & 07:42:45 \\
			Modaresi \& Gross & 0.350 & 0.603 & 0.580 & 0 & 00:00:07 \\
			Khonji \& Iraqi & 0.349 & 0.599 & 0.583 & 1 & 09:10:01 \\
			Halvani \& Steinebach & 0.338 & 0.629 & 0.538 & 1 & 00:00:07 \\
			Zamani et al. & 0.322 & 0.585 & 0.550 & 0 & 00:02:03 \\
			Mayor et al. & 0.318 & 0.572 & 0.557 & 10 & 01:01:07 \\
			Castillo et al. & 0.318 & 0.549 & 0.580 & 0 & 01:31:53 \\
			Harvey & 0.312 & 0.579 & 0.540 & 0 & 00:10:22 \\\hline
			\textbf{Our approach} & \textbf{0.312} & \textbf{0.558} & \textbf{0.560} & \textbf{0} & \textbf{00:00:05} \\\hline
			{\color{gray} BASELINE} & {\color{gray} 0.288} & {\color{gray} 0.543} & {\color{gray} 0.530} & {\color{gray} 0} & {\color{gray} 00:03:29} \\
			Jankowska et al. & 0.284 & 0.518 & 0.548 & 5 & 01:16:35 \\
			Vartapetiance \& Gillam & 0.270 & 0.520 & 0.520 & 0 & 00:16:44 \\ \hline
	\end{tabular}}
	\caption{PAN 2014 results on the English essays evaluation corpus. The final score FS denotes \auccat and UP the number of unanswered problems. Table adapted from \cite{PAN14Stamatatos:2014}.}
	\label{tab:PAN2014esEval}
\end{table}

\begin{table}
	\centering
	\scalebox{0.84}{
		\begin{tabular}{|l|r|r|r|r|r|}
			\hline
			\bf Team & \bf FS & \bf AUC & \bf c@1  & \bf UP  & \bf Runtime \\ \hline
			Modaresi \& Gross & 0.508 & 0.711 & 0.715 & 0 & 00:00:07 \\
			Zamani et al. & 0.476 & 0.733 & 0.650 & 0 & 02:02:02 \\
			{\color{gray} META-CLASSIFIER} & {\color{gray} 0.472} & {\color{gray} 0.732} & {\color{gray} 0.645} & {\color{gray} 0} &  \\
			Khonji \& Iraqi & 0.458 & 0.750 & 0.610 & 0 & 02:06:16 \\\hline
			\textbf{Our approach} & \textbf{0.421} & \textbf{0.695} & \textbf{0.605} & \textbf{0} & \textbf{00:00:14} \\\hline
			Mayor et al. & 0.407 & 0.664 & 0.614 & 8 & 01:59:47 \\
			Castillo et al. & 0.386 & 0.628 & 0.615 & 0 & 02:14:11 \\
			Satyam et al. & 0.380 & 0.657 & 0.579 & 3 & 02:14:28 \\
			Frery et al. & 0.360 & 0.612 & 0.588 & 1 & 00:03:11 \\
			Moreau et al. & 0.313 & 0.597 & 0.525 & 12 & 00:11:04 \\
			Halvani \& Steinebach & 0.293 & 0.569 & 0.515 & 0 & 00:00:07 \\
			Harvey & 0.283 & 0.540 & 0.525 & 0 & 00:46:30 \\
			Layton & 0.260 & 0.510 & 0.510 & 0 & 07:27:58 \\
			Vartapetiance \& Gillam & 0.245 & 0.495 & 0.495 & 0 & 00:13:03 \\
			Jankowska et al. & 0.225 & 0.491 & 0.457 & 1 & 02:36:12 \\
			{\color{gray} BASELINE} & {\color{gray} 0.202} & {\color{gray} 0.453} & {\color{gray} 0.445} & {\color{gray} 0} & {\color{gray} 00:08:31} \\ \hline
	\end{tabular}}
	\caption{PAN 2014 results on the English novels evaluation corpus. The final score FS denotes \auccat and UP the number of unanswered problems. Table adapted from \cite{PAN14Stamatatos:2014}.}
	\label{tab:PAN2014noEval}
\end{table}

\begin{table}
	\centering
	\scalebox{0.84}{
		\begin{tabular}{|l|r|r|r|r|r|}
			\hline
			\textbf{Submission} & $\bm{F_1}$ & \textbf{Precision} & \textbf{Recall} \\ \hline
			Veenman\&Li & 0.800 & 0.800 & 0.800 \\\hline
			\textbf{Our approach} & \textbf{0.800} & \textbf{0.800} & \textbf{0.800} \\\hline
			Seidman & 0.800 & 0.800 & 0.800 \\
			Layton et al. & 0.767 & 0.767 & 0.767 \\
			Moreau\&Vogel & 0.767 & 0.767 & 0.767 \\
			Jankowska et al. & 0.733 & 0.733 & 0.733 \\
			Vilariño et al. & 0.733 & 0.733 & 0.733 \\
			Halvani et al.  & 0.700 & 0.700 & 0.700 \\
			Feng\&Hirst & 0.700 & 0.700 & 0.700 \\
			Ghaeini  & 0.691 & 0.760 & 0.633 \\
			Petmanson  & 0.667 & 0.667 & 0.667 \\
			Bobicev  & 0.644 & 0.655 & 0.633 \\
			Sorin & 0.633 & 0.633 & 0.633 \\
			van Dam  & 0.600 & 0.600 & 0.600 \\
			Jayapal\&Goswami  & 0.600 & 0.600 & 0.600  \\
			Kern  & 0.533 & 0.533 & 0.533 \\
			{\color{gray} BASELINE} & {\color{gray} 0.500} & {\color{gray} 0.500} & {\color{gray} 0.500} \\
			Vartapetiance\&Gillam  & 0.500 & 0.500 & 0.500  \\
			Ledesma et al.  & 0.467 & 0.467 & 0.467  \\
			Grozea & 0.400 & 0.400 & 0.400 \\ \hline
	\end{tabular}}
	\caption{PAN 2013 results (in terms of $F_{1}$) on the English evaluation corpus. Table adapted from \cite{PAN13Stamatatos:2013}.}
	\label{tab:PAN2013EvalAccuracy}
\end{table}

\begin{table}
	\centering
	\scalebox{0.84}{
		\begin{tabular}{|l|l|r|}
			\hline
			\textbf{Rank} & \textbf{Submission} & \textbf{AUC} \\ \hline
			1 & Jankowska, et al. & 0.842 \\
			2 & Ghaeini & 0.837 \\\hline
			3 & \textbf{Our approach} & \textbf{0.813}  \\\hline
			4 & Seidman & 0.792 \\
			5 & Feng\&Hirst & 0.750 \\
			6 & Petmanson & 0.672 \\
			7 & Bobicev & 0.585 \\
			{\color{gray} -} & {\color{gray} BASELINE} & {\color{gray} 0.500} \\
			8 & Kern & 0.384 \\
			9 & Grozea & 0.342 \\
			10 & Layton et al. & 0.277 \\ \hline
	\end{tabular}}
	\caption{PAN 2013 results (in terms of AUC) on the English evaluation corpus. Table adapted from \cite{PAN13Stamatatos:2013}.}
	\label{tab:PAN2013EvalAUC}
\end{table}

\subsection{Experiment 3: Evaluation (baselines)} 
As we have seen in Experiment 2, our \AV method performed quite well regarding three out of four PAN evaluation corpora. The goal of the following experiment is to test our method against three baselines (mentioned in Section~\ref{RelatedWork}) on all seven evaluation corpora (described in Section~\ref{Corpora}). From the results, listed in Table~\ref{tab:HalvaniVsBaselines}, we can see that GLAD outperforms our method as well as the other baselines in four cases. On the other hand, our method outperformed all baselines in two cases and achieved, in comparison to GLAD, similar results regarding $\panThirteenEval$ and $\panFourteenNoEval$, in terms of \auccat (or just AUC). 

We can infer from Table~\ref{tab:HalvaniVsBaselines} that the $\panFourteenEsEval$ corpus is quite challenging, not only for our method but also for the \StamatatosMethod and \KoppelMethod. GLAD, in contrast, achieves good results on the same corpus. In fact, a closer look on Table~\ref{tab:PAN2014esEval} reveals that GLAD outperforms all participated \AV approaches (except the meta-model) which indicates its good generalization ability. A similar picture can be seen regarding $\AmazonEval$, where GLAD not only significantly outperformed our method and the two baselines, but also achieved the highest AUC (0.937) across all methods on all seven corpora. Surprisingly, $\panFifteenEval$ is the only corpus where GLAD yielded the lowest recognition result. We assume that the characteristic of $\panFifteenEval$ (one known vs. one unknown document with an average number of 536 word per document \cite{PAN15Stamatatos:2015}) causes the moderate performance of GLAD on it. 

Another interesting observation that can be seen in Table~\ref{tab:HalvaniVsBaselines} (AUC column) is that in most cases all involved methods perform better on the three additional corpora than on the four PAN evaluation corpora. In particular, the $\KoppelEval$ corpus seems not to be challenging for all methods as they achieved here the highest results from all seven corpora. 

\begin{table*}
	\small
	\centering
	\resizebox{\textwidth}{!}{
		\begin{tabular}{|l||r|r|r|r||r|r|r|r||r|r|r|r|}
			\hline
& \multicolumn{4}{c||}{\textbf{\auccat}} & \multicolumn{4}{c||}{\textbf{AUC}}  & \multicolumn{4}{c|}{\textbf{c@1}} \\ \hline
Corpus & $\AVhalvani$ & $\AVstamatatos$ & $\AVglad$ & $\AVkoppel$ & $\AVhalvani$ & $\AVstamatatos$ & $\AVglad$ & $\AVkoppel$ & $\AVhalvani$ & $\AVstamatatos$ & $\AVglad$ & $\AVkoppel$ \\\hline
$\panThirteenEval$   & 0.650 & 0.461 & 0.664 & \textbf{0.694} & 0.813 & 0.692 & \textbf{0.866} & 0.833 & 0.800 & 0.667 & 0.767 & \textbf{0.833} \\
$\panFourteenEsEval$ & 0.312 & 0.339 & \textbf{0.514} & 0.289 & 0.558 & 0.599 & \textbf{0.751} & 0.540 & 0.560 & 0.565 & \textbf{0.685} & 0.535 \\
$\panFourteenNoEval$ & 0.421 & 0.397 & \textbf{0.449} & 0.354 & 0.695 & 0.702 & \textbf{0.709} & 0.599 & 0.605 & 0.565 & \textbf{0.633} & 0.590 \\
$\panFifteenEval$    & \textbf{0.605} & 0.393 & 0.406 & 0.254 & \textbf{0.802} & 0.719 & 0.644 & 0.504 & \textbf{0.754} & 0.546 & 0.630 & 0.504 \\
$\RedditEval$        & 0.529 & 0.549 & \textbf{0.681} & 0.471 & 0.748 & 0.773 & \textbf{0.868} & 0.698 & 0.708 & 0.710 & \textbf{0.784} & 0.675 \\
$\KoppelEval$        & \textbf{0.789} & 0.646 & 0.764 & 0.589 & \textbf{0.926} & 0.847 & 0.912 & 0.792 & \textbf{0.852} & 0.763 & 0.838 & 0.744 \\
$\AmazonEval$        & 0.631 & 0.577 & \textbf{0.817} & 0.632 & 0.842 & 0.793 & \textbf{0.937} & 0.831 & 0.749 & 0.728 & \textbf{0.872} & 0.760 \\\hline \hline
Average				 & 0.562 & 0.480 & \textbf{0.614} & 0.469 & 0.769 & 0.732 & \textbf{0.812} & 0.685 & 0.718 & 0.649 & \textbf{0.744} & 0.663 \\\hline
		\end{tabular}
	}
	\caption{Our approach ($\AVhalvani$) vs. \StamatatosMethod ($\AVstamatatos$), \GLADMethod ($\AVglad$) and \KoppelMethod ($\AVkoppel$).}
	\label{tab:HalvaniVsBaselines}
\end{table*}


\section{Conclusions} \label{Conclusions} 
We presented a simple but quite effective and efficient \av method, based on compression models. The method yields competitive results compared to current state-of-the-art approaches, that are based on sophisticated machine learning models such as support vector machines or neural networks. In contrast to these approaches, our \AV method only relies on a compressing algorithm, a simple dissimilarity measure and a threshold, acting as an acceptance criterion regarding the authorship of a questioned document. Moreover, our method does not require natural language processing concepts (\eg part-of-speech tagging or parsing) or even less complex techniques such as regular expressions, which are used in many traditional \AV approaches. 

In a series of experiments we made a number of observations regarding our approach. In the first experiment we have shown that PPMd was able to outperform all other tested compressors. In addition, the CBC measure yielded the highest results, which behaved stable across the five compressors that were tested against all training corpora. In the second experiment we compared our method against a fair number of \AV approaches, which were submitted to three international \AV competitions. With an exception of one out of four evaluation corpora, our method performs amongst the highest-ranked \AV methods. In the same experiment we provided the runtime of our method for comparison, where it turned out that it counts to the fastest approaches. In the third experiment we evaluated our method against three strong baselines on seven evaluation corpora and achieved competitive verification results. 

However, despite of the success and the benefits of our \AV method, a number of questions remain open and must be addressed in future work. One important issue, for example, that was not taken into account in our study is a deeper investigation regarding the proposed method. In fact, at this stage, we cannot provide a satisfactory explanation regarding the good performance of our method and which concrete limitations it actually faces. Our experiments revealed that the method was able to solve even complicated \AV tasks such as cross-genre/topic problems which, in addition, were short in terms of text lengths. Hence, we do not consider these as limitations for the method. Besides this, robustness must be covered in future work, including under which conditions the method fails to solve verification problems, which it solved beforehand. And last but not least, we plan to construct an ensemble based on our method and other successful methods (for example GLAD) in order to achieve strong and reliable verification results.  
\newpage
\bibliographystyle{plain}
\bibliography{references}
\end{document}